\documentclass[12pt]{article}
\usepackage{epsfig}

\usepackage{a4}

\usepackage{amsmath}

\begin{document}
\title{Regular and black hole solutions to higher order curvature
Einstein--Yang-Mills--Grassmannian systems in $5$ dimensions}
\author{{\large Yves Brihaye,}$^{\ddagger}$
{\large Eugen Radu}$^{\dagger}$
and {\large D. H. Tchrakian}$^{\dagger \star}$ \\ \\
$^{\ddagger}${\small Physique-Math\'ematique, Universite de
Mons-Hainaut, Mons, Belgium}\\ \\
$^{\dagger}${\small Department of
Mathematical Physics, National University of Ireland Maynooth,} \\
{\small Maynooth, Ireland} \\
$^{\star}${\small School of Theoretical Physics -- DIAS, 10 Burlington
Road, Dublin 4, Ireland }}

\date{}
\newcommand{\dd}{\mbox{d}}
\newcommand{\tr}{\mbox{tr}}
\newcommand{\la}{\lambda}
\newcommand{\ka}{\kappa}
\newcommand{\al}{\alpha}
\newcommand{\ga}{\gamma}
\newcommand{\de}{\delta}
\newcommand{\si}{\sigma}
\newcommand{\bomega}{\mbox{\boldmath $\omega$}}
\newcommand{\bsi}{\mbox{\boldmath $\sigma$}}
\newcommand{\bchi}{\mbox{\boldmath $\chi$}}
\newcommand{\bal}{\mbox{\boldmath $\alpha$}}
\newcommand{\bpsi}{\mbox{\boldmath $\psi$}}
\newcommand{\brho}{\mbox{\boldmath $\varrho$}}
\newcommand{\beps}{\mbox{\boldmath $\varepsilon$}}
\newcommand{\bxi}{\mbox{\boldmath $\xi$}}
\newcommand{\bbeta}{\mbox{\boldmath $\beta$}}
\newcommand{\ee}{\end{equation}}
\newcommand{\eea}{\end{eqnarray}}
\newcommand{\be}{\begin{equation}}
\newcommand{\bea}{\begin{eqnarray}}
\newcommand{\ii}{\mbox{i}}
\newcommand{\e}{\mbox{e}}
\newcommand{\pa}{\partial}
\newcommand{\Om}{\Omega}
\newcommand{\vep}{\varepsilon}
\newcommand{\bfph}{{\bf \phi}}
\newcommand{\lm}{\lambda}
\def\theequation{\arabic{equation}}
\renewcommand{\thefootnote}{\fnsymbol{footnote}}
\newcommand{\re}[1]{(\ref{#1})}
\newcommand{\R}{{\rm I \hspace{-0.52ex} R}}
\newcommand{\N}{{\sf N\hspace*{-1.0ex}\rule{0.15ex}%
{1.3ex}\hspace*{1.0ex}}}
\newcommand{\Q}{{\sf Q\hspace*{-1.1ex}\rule{0.15ex}%
{1.5ex}\hspace*{1.1ex}}}
\newcommand{\C}{{\sf C\hspace*{-0.9ex}\rule{0.15ex}%
{1.3ex}\hspace*{0.9ex}}}
\newcommand{\eins}{1\hspace{-0.56ex}{\rm I}}
\renewcommand{\thefootnote}{\arabic{footnote}}

\maketitle


\bigskip

\begin{abstract}
Solutions to EYM systems in $5$ spacetime dimensions possessing no
gravity decoupling limits, feature a
peculiar critical behaviour which is absent in their $6,7$ and $8$
dimensional counterparts which do possess flat space limits. This
critical behaviour in $5$ dimensions
persists even when a scalar matter field is added, rendering the model
nontrivial in the gravity decoupling limit. To this end, both
regular and black hole spherically symmetric solutions to the
higher curvature EYM--Grassmannian sigma model model in $d=5$ spacetime
dimensions are constructed. A study of the solutions
to the Grassmannian model in flat space is also carried out.
\end{abstract}
\medskip
\medskip
\newpage

\section{Introduction}
\label{introduction}
Gravitational theories in higher dimensions are of current interest in
the contexts of (10 dimensional) superstring theory and of theories
with large and infinite extra dimensions with non-factorisable metrics.
It is thus interesting to study the properties of the
corresponding Einstein--Yang-Mills (EYM) systems in higher dimensions
generalising the usual four dimensional EYM model whose regular solutions
were constructed in \cite{BK}, and black-hole solutions in
\cite{black1,black2}.

Since the low energy effective action of string theory includes
higher order terms in the gravitational and Yang-Mills (YM) curvatures,
such systems were recently studied in spacetime dimensions
$d=6,7,8$~\cite{bct2} and $d=5$~\cite{bcht}, in which dimensions it is
necessary to include higher curvature YM terms to enable the existence
of particle like solutions.

As expected, the gravitating YM models in spacetime dimensions $d=5,6,7,8$
support particle like solutions for a finite range of the gravitational
parameter $\alpha^2$, say to $\alpha_{max}^2$. This is similar to the
case of gravitating monopoles~\cite{bfm,w} in $d=4$. What actually happens
here differently from the latter case
is that two solutions exist for a given value of $\alpha^2$. But
there is a marked difference between the $d=6,7,8$ cases~\cite{bct2},
where there exist two solutions for all values of $\alpha^2$, and
the $d=5$ case~\cite{bcht} where this is true for values of $\alpha^2$
up to a critical value $\alpha_{c}^2$. In the latter case, solutions
exist for values of $\alpha^2$ oscillating about $\alpha_{c}^2$. The
tracking of this peculiar singular behaviour in the $d=5$ model is
the aim of the present work.

Now the most obvious difference between the solutions of the
models~\cite{bct2} in $d=6,7,8$ on the one hand, and those of the
model~\cite{bcht} in $d=5$ on the other, is that the solutions in the
former cases persist in the gravity decoupling limit~\footnote{The
solutions in the flat space limit of the $d=6,7,8$ models~\cite{bct2}
are exemplified by the instanton--like solution studied in \cite{butc}.},
while those in the $d=5$ model do not~\footnote{Note
that due the Derrick scaling requirement the YM field in the static $4$
Euclidean dimensions supports a 'soliton' only if the YM system consists
of the $p=1$ term, exclusively. On the other hand when gravity is
switched on, the absence of the $p=2$ YM term (see Ref. \cite{T} for the
YM hierarchy) prevents the existence of a
soliton, due to the same scaling requirement.}. We have therefore added
a scalar matter field to
the $d=5$ model, which renders the gravity decoupling limit nontrivial,
to investigate the nature of the critical behaviour discovered in
\cite{bcht}. We thus answer the question: Is this critical behaviour a
consequence of the absence of a gravity decoupling limit in $d=5$,
or is it a peculiarity of the dimensionality of the spacetime itself? Our
answer is, that this property pertains to the dimensionality of the
spacetime, since we find that it persists also in the new model which
supports a gravity decoupling solution.

For this purpose we introduce an $SU(2)$ gauged $4\times 2$ Grassmannian
field~\cite{MCT} describing a sigma model, to the $d=5$ EYM model. The
introduction of this scalar field is analogous
to the inclusion of a Higgs\footnote{The choice of a Grassmannian field
rather than a Higgs field is because
the gauge connection in a $d-1$ dimensional Higgs model~\cite{OAT}
supporting a 'soliton' behaves as {\it one half pure gauge}
asymptotically, resulting in $r^{-1}$ decay. As a result the integral of
the $p=1$ YM term is convergent only in $(d-1)\le 3$ so that only higher
$p$ YM terms are admissible for $(d-1)\ge 4$. So if we insist in keeping
the usual $p=1$ YM term, then we must avoid using of a (generalised) Higgs
model~\cite{OAT}.  This contrasts with the faster decay of an 'instanton'
in a pure gauge theory where the gauge connection is {\it pure gauge} and
hence the integral of the $p=1$ YM term is convergent in $d-1=4$. It
turns out that the connection of the gauged Grassmannian model~\cite{MCT}
is asymptotically {\it pure gauge}, and hence our choice.} field in $d=4$
as in Refs.~\cite{bfm,w}.

We find that the singular behaviour in question, which is peculiar to the
solutions in $d=5$ spacetime only, presists whether or not the model
supports a regular solution in the flat space limit. 

\section{The model and the equations}
\label{modelsandeqns}
In the first subsection we give the Lagrangian of our $4+1$ dimensional
model in Minkowski space, and in the second one we impose static
spherically symmetry and write down the resulting one dimensional
ordinary differential equations.
\subsection{The models}
\label{models}

We take the gravitational and YM sectors of the model to be precisely
the one analysed in \cite{bcht}, augmented by the Grassmannian sigma
model term
\be
\label{L}
{\cal L}={\cal L}_{grav}\ +\ {\cal L}_{YM}\ +\ {\cal L}_{grass}\ ,
\ee
in $5$ spacetime dimensions. The first two terms of \re{L} describe
the EYM sector defined as
\be
\label{gravYM}
{\cal L}_{grav}=e\ \frac{\ka_1}{2}\ R_{(1)}\quad ,\quad
{\cal L}_{YM}=e\ \left(\frac{\tau_1}{4}\ {\mbox Tr\ }F(2)^2+
\frac{\tau_2}{48}\ {\mbox Tr\ }F(4)^2\right)\ ,
\ee
where $R_{(1)}$ describes the usual Einstein gravity, and
$e=\mbox{det}\ e_{\mu}^a=\sqrt{-\mbox{det}\,g_{\mu\nu}}$,
$e_{\mu}^a$ are the $5$-beins.
$F(2p)$, for $p=1,2$, is the $2p$-form YM curvature (see Refs.
\cite{bct2} and \cite{T}), which for $p=1$ is the usual $2$-form YM
curvature taking values in the antihermtian representation of the
algebra of $SU(2)$ here.

The Grassmannian sigma model part of \re{L} is
\be
\label{grass}
{\cal L}_{grass}=e\,\mbox{Tr} \left(\tau D_{\mu}z^{\dagger}D^{\mu}z
+\frac12\tau\mu^2(\eins+z^{\dagger}\Gamma_5z)
+\la(\eins-z^{\dagger}z)\right)
\ee
described by the $4\times 2$ Grassmannian field subject to the $2\times 2$
condition
\[
z^{\dagger}z=\eins
\]
and whose $SU(2)$ covariant derivative is defined by
\be
\label{cov}
D_{\mu}z=\pa_{\mu}z-z\, A_{\mu}\ .
\ee
In \re{grass} the constant $\mu$ has the dimensions of mass and leads to
the exponential localisation of the ensuing topologically stable lump,
and the Lagrange multiplier $\la$ is a $2\times 2$ array. The gravity
decoupled version of this model is the gauged Grassmannian sigma model
for which the 'instanton' solutions were constructed numerically in
\cite{BBT}, which will be studied in more detail here.

\subsection{The classical equations}
\label{classeqns}
In $d$ dimensional spacetime, we restrict to static fields that
are spherically symmetric in the $d-1$ spacelike dimensions with the
metric Ansatz
\be
\label{metric}
ds^2=-\si(r)^2N(r)dt^2+N(r)^{-1}dr^2+r^2d\Omega_{d-2}^2
\ee
where $r$ is the spacelike radial coordinate and
$d\Omega_{d-2}$ is the $d-2$ dimensional angular volume element.

We take the static spherically symmetric $SU(2)$ YM field in $5$
spacetime (i.e. $4$ Euclidean) dimensions, in one or other
chiral representation of $SO_{\pm}(4)$, to be
\be
\label{YMsph}
A_0=0\ ,\quad
A_i=\left(\frac{1-w}{r}\right)\Sigma_{ij}^{(\pm)}\hat x_j\ , \quad
\Sigma_{ij}^{(\pm)}=-\frac{1}{4}\left(\frac{1\pm\gamma_5}{2}\right)
[\gamma_i ,\gamma_j]\ ,
\ee
where
\[
\Sigma_{ij}^{(\pm)}=
-\frac{1}{4}\Sigma^{(\pm)}_{[i}\,\Sigma^{(\pm)}_{j]}
\]
where in a more familiar notation $\Sigma^{(+)}_i=\si_i=
(i\vec\si,\eins)$ and $\Sigma^{(-)}_i=\tilde\si_i=
(-i\vec\si,\eins)$, with $i=1,2,3,4$ and in terms of the three Pauli spin
matrices $\vec\si$.

The spherically symmetric Ansatz for the Grassmannian field $z$, whose
consistency has been checked, is
\be
z=\left[
\begin{array}{c}
\sin\frac{f}{2}\,\ \ \ \ \eins \\
\cos\frac{f}{2}\,\,\hat x_i\tilde\si_i  \\
\end{array}
\right]\ .
\label{grasssph}
\ee
Subjecting \re{gravYM} and \re{grass} to spherical
symmetry by empoloying the
Ans\"atze \re{metric}, \re{YMsph} and \re{grasssph}, and subjecting the
resulting one dimensional Lagrange density to the variational principle,
we find the following equations for the functions $f(r)$, $w(r)$, $N(r)$
and $\si(r)$,

\medskip
\be
\label{grasseq}
(r^3\si Nf')'+r\, \si\left(3w-\mu^2r^2\right)\sin f=0,
\ee

\medskip
\be
\tau_1\left(\left(r\si Nw'\right)'-2r^{-1}\si(w^2-1)w\right)
+3\tau_2(w^2-1)\left(r^{-3}\si N(w^2-1)w'\right)'
=2\tau\, r\, \si(w+\cos f),
\label{YM12eq}
\ee

\medskip
\bea
m'&=&\frac{1}{8}r\left(\tau_1\left[Nw'^2
+\left(\frac{w^2-1}{r}\right)^2\right]
+\frac{3}{r^2}\tau_2\left(\frac{w^2-1}{r}\right)^2Nw'^2\right)\nonumber\\
&+&\frac{\tau}{12}r^3\left[Nf'^2+\frac{3}{r^2}\left(w^2+2w\cos f+1\right)
+2\mu^2(1-\cos f)\right],
\label{meq}
\eea

\medskip
\be
\ka_1\left(
\frac{\si'}{\si}\right)=\frac{n_5}{8r}
\left[\tau_1+\frac{3}{r^2}\tau_2
\left(\frac{w^2-1}{r}\right)^2\right]w'^2
+\frac{n_5}{12}\tau\, rf'^2\ .
\label{sigeq}
\ee
The ADM mass $M=\lim_{r\to\infty}m(r)$, with $m(r)$ defined as
\be
\label{mdef}
m(r)=n_5^{-1}\ka_1r^2(1-N)\ .
\ee
For $N(r)=\si(r)=1$ everywhere, namely the gravity decoupling limit,
\re{grasseq} and \re{YM12eq} satisfy the flat space solution.

In the numerical work below, we set the self interaction potential of the
Grassmannian field, $\mu^2=0$ without changing the qualitative nature of
the solutions. This is because in this model, as in
Higgs models, the finite energy condition leads to a unique asymptotic
value for the matter field, as can be seen easily by inspection of the
term multiplying $\tau$ in \re{meq}. This situation contrasts with that
in certain other gauged nonlinear sigma models, where a unique asymptotic
value for the matter field can be ensured only by the inclusion of a self
interaction potential, e.g. the pion mass potential of the Skyrme model,
leading to bifurcations~\cite{BT}.

\subsection{Boundary values and asymptotic behaviour}
\label{boundary}
In the next section, we will solve the above equations with the
appropriate boundary
conditions for the radial functions $m(r)$, $\si(r)$ $w(r)$ and $f(r)$
which guarantee the solution to be regular at the origin and to have
finite energy.
For regular solutions, the boundary conditions at the origin are
\be
\label{origin}
m(0)=0\ \ ,\qquad w(0)=1,\qquad f(0)=\pi,
\ee
while the conditions satisfied on the event horizon $r=r_h$ are
\begin{eqnarray}
\label{eh-g}
N(r_h)=0,~~~\sigma(r_h)=\sigma_h,~~~
w(r_h)=w_h,~~~ f(r_h)=f_h,
\end{eqnarray}
with $\sigma_h,w_h,f_h$ real constants.
The asymptotic form of the solution is
\be
\label{infty}
\lim_{r\to\infty}\si(r)=1\ \ ,\qquad \lim_{r\to\infty}w(r)=-1\qquad,
\ \lim_{r\to\infty}f(r)=0.
\ee
The condition on $\si(r)$ results in the metric being
asymptotically Minkowskian. 

The asymptotic solutions to these functions can be systematically
constructed in both regions, near the origin (or event horizon)
and for $ r \gg 1$. Defining $\alpha^2 = \frac{n_d}{8 \kappa_1}$
we find for $r \ll 1$
\begin{eqnarray}
\label{sig0}
&f(r)&=\pi-c_3r+o(r^3)\ \ , \nonumber \\
&w(r) &= 1 + c_1 r^{2} + o(r^{4})\  \ \ , \nonumber \\
&\sigma(r) &= \sigma_0[1+2\alpha^2 c_1^2 r^2
(\tau_1 + 12\tau_2c_1^2]+o(r^4)) \ \ ,  \nonumber  \\
&m(r) &= \frac{1}{4} r^4 [c_1^2 (\tau_1 + 6 c_1^2 \tau_2)+\frac16\tau
c_3^2]+ o(r^{6}).
\end{eqnarray}
For black hole configurations, the expression of the solutions
near the event horizon is
\begin{eqnarray}
\label{eh}
\nonumber
f(r)&=&f_h+f'(r_h)
(r-r_h)+O((r-r_h)^2),
\nonumber
\\
w(r)&=&w_h+w'(r_h)
(r-r_h)+O((r-r_h)^2),
\nonumber
\\
\sigma (r)&=& \sigma_h \alpha^2
\left(1+\Big(
(\tau_1+\frac{3\tau_2(w_h^2-1)^2}{r_h^4}) w'^2(r_h)
+\frac{2}{3}\tau r_h f'^{2}(r_h) \Big)
(r-r_h) \right)
+O((r-r_h)^2),\nonumber
\\
m(r)&=& \frac{r_h^2}{8 \alpha^2}+m'(r_h)
(r-r_h)+O((r-r_h)^2),
\end{eqnarray}
where
\begin{eqnarray}
\nonumber
m'(r_h)&=&\frac{\tau_1(w_h^2-1)^2}{8r_h}
+\frac{\tau r_h^3}{12}\left(\frac{3}{r_h^2}(w_h^2+2w_h \cos f_h+1)
+2\mu^2(1-\cos f_h)\right),
\\
\nonumber
w'(r_h)&=&\frac{2\tau r_h(w_h+\cos f_h)+2\tau_1 {w_h(w_h-1)}/{r_h}}
{(\tau_1r_h+3\tau_2 {(w_h^2-1)^2}/{r_h^3})
(- 8\alpha^2m'(r_h)/r_h^2 + {2}/{r_h})}\,,\nonumber\\
f'(r_h)&=&-\frac{r_h (3w_h^2-\mu^2 r_h^2)\sin f_h}
{r_h^3(- 8\alpha^2m'(r_h)/r_h^2 + {2}/{r_h})}\,.
\end{eqnarray}
For $r \gg 1$ we find
\begin{eqnarray}
\label{siginf}
&f(r)&=r^{-1}K_2(\mu r) \ \ , \nonumber \\
&w(r) &= -1 + \frac{c_2} {r^{d-3}} \  \ \ , \nonumber \\
&\sigma(r) &=1 - \frac {\tau_1 \alpha^2 c_2^2 (d-3)^2 }
{(2d-4)  r^{2d-4}}
 \ \ ,  \nonumber  \\
&m(r) &= m_{\infty} - \frac{\tau_1 (d-3) c_2^2}{8 r^{d-1}}
\end{eqnarray}
In the first member of \re{siginf}, $K_2(\mu r)$ is a Bessel function
leading to exponential decay by virtue of the mass term $\mu$ in
the potential.
The constants $c_1, c_2 , c_3 , \sigma_0, f_h,w_h,\sigma_h
 $ and $m_{\infty}$ have 
to be determined numerically; $m_{\infty}$ is nothing
else but the ADM mass of the solution as noted previously.
They depend generically on the coupling constants of the theory.

\section{Numerical results}
\label{numerics}
In this section we will present three sets of results. These are the
non gravitating $SU(2)$ gauged Grassmannian solitons, the gravitating
regular solutions of this system, and, the corresponding black hole
solutions. Each will be reported in a separate subsection.

The numerical integrations were carried out using both a shooting method
as well as applying the numerical program COLSYS~\cite{colsys}, with
complete agreement to very high accuracy.
\subsection{$SU(2)$ gauged Grassmannian solitons}
\label{gauged Grass}
Since an important reason for considering the model in this work is
that it supports a non gravitating flat space limit, it is worth studying
the properties of the latter for its own sake. This is especially so since
we find that this (non gravitating) model supports solutions which
exhibit nodes in the profile of the Grassmann function $f(r)$ defined in
\re{grasssph}.
\newpage
\setlength{\unitlength}{1cm}
\begin{picture}(18,8)
\centering
\put(1,0){\epsfig{file=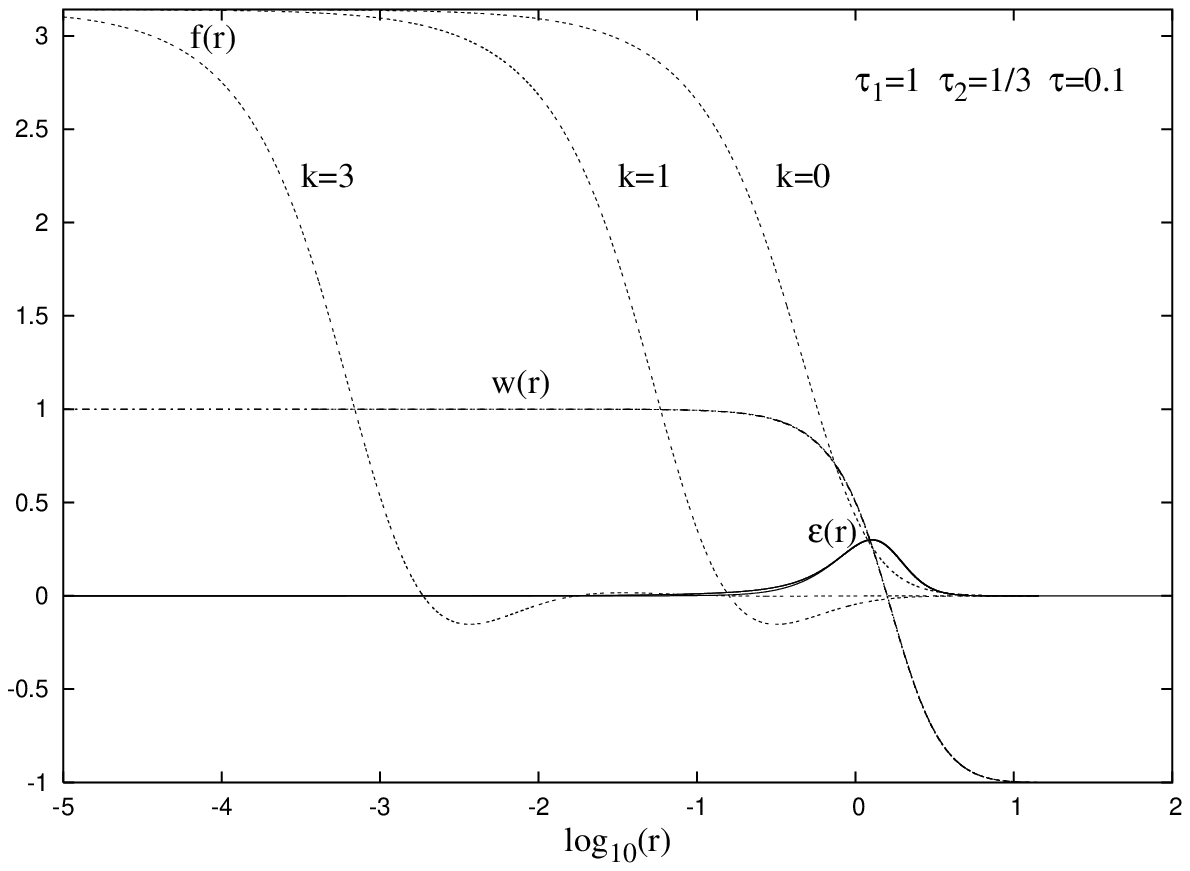,width=14cm}}
\end{picture}
\\
\\
{\small {\bf Figure 1.}
The functions $w(r)$ and $f(r)$ and the energy density
$\epsilon(r)$
are plotted as functions of radius
for typical flat space solutions with the coupling
constants $\tau_1=3\tau_2=1,\tau=0.1$.
The node number $k$ of the Grassmanian function $f(r)$ is also indicated}.
\\
\\
 The profiles of the YM functions $w(r)$ do not change
appreciably for the solutions with different number of nodes of $f(r)$,
neither qualitatively nor quantitatively, exhibiting only one node.
Also, somewhat unexpected,
the total mass/energy of these solutions stays  almost constant, when
increasing the node number of $f(r)$ (with differences less than one
percent). We shall see below that all the multinode solutions result in
qualitatively very similar gravitating solutions, which we will exploit
to simplify the numerical work section \re{Blackhole}.

It turns out that there are solutions with infinitely many nodes in the
function $f(r)$, such that the profiles contract towards the origin as
the number of nodes $n$ increases. The profiles of $w(r)$ and the energy
density remain insensitive to the increase in the number of nodes $n$
of $f(r)$, as seen in Figure 1. In the limit $n\to\infty$ the profile of
$f(r)\to f_{\infty}(r)$ shrinks to the origin such that
\[
f_{\infty}(r)=0 \qquad {\rm for\ \ all}\ \ r\,.
\]
We refer to this configuration as that of frozen $f$, namely $f=0$
everywhere. Also, for all considered solutions, as well 
as the gravitating counterparts, we find that the gauge function $w(r)$
monotonically decreases towards its asymptotic value
without presenting local extrema.

\subsection{Regular gravitating solutions}
\label{Regular}
We numerically integrate the Eqs. (\ref{grasseq})-(\ref{sigeq})
with the boundary conditions
(\ref{origin}), (\ref{infty}) for 
$\tau_1=1,\tau_2=1/3$ and several values of $\tau$,
finding the following picture.
First, for $\alpha^2$ being small enough, a branch of
solutions smoothly emerges from the flat space configurations.
When $\alpha^2$ increases,  the mass parameter $M$ decreases, as well as
the value $\sigma(0)$ and the minimum $N_m$
\newpage
\setlength{\unitlength}{1cm}
\begin{picture}(18,8)
\centering
\put(1,0){\epsfig{file=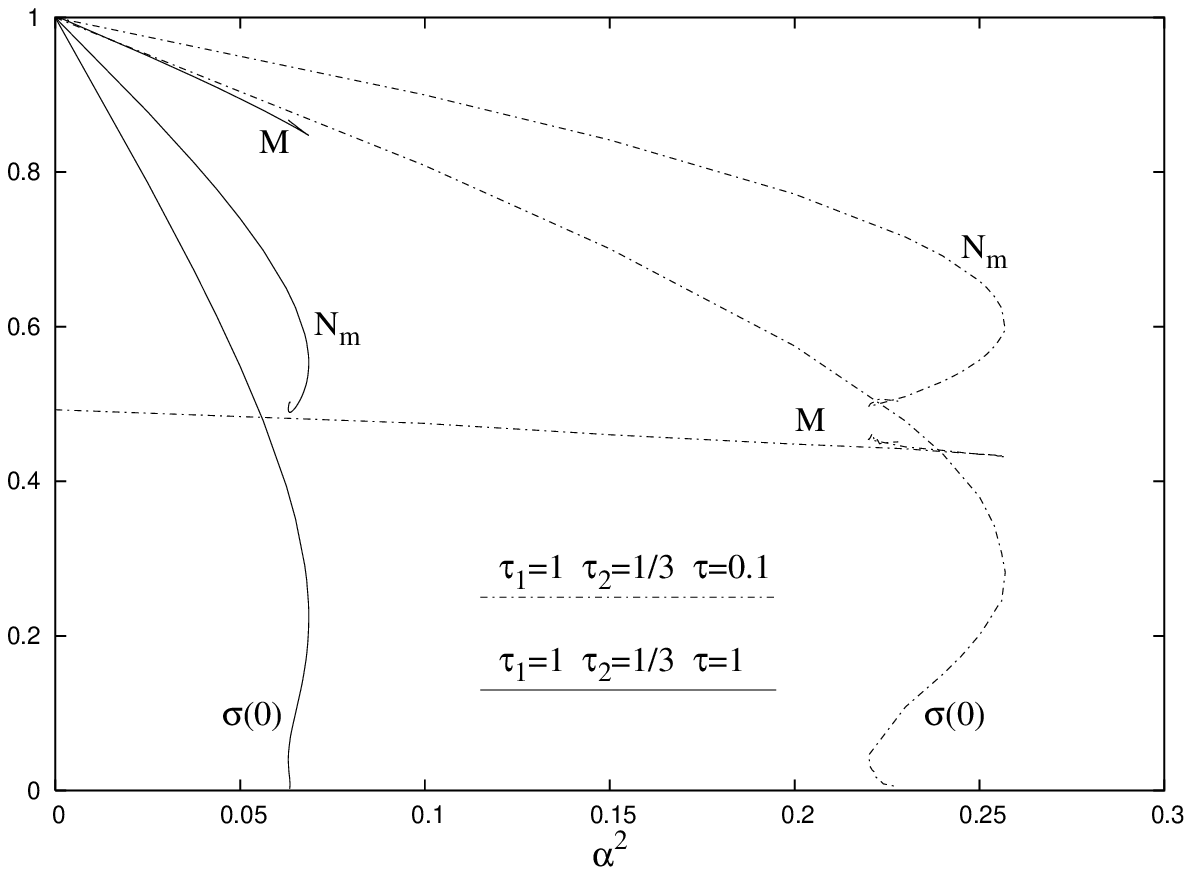,width=14cm}}
\end{picture}
\\
\\
{\small {\bf Figure 2.}
The value $N_m$ of the minimum of the metric function $N$,
the value of the metric function $\sigma$ at the origin $\sigma(0)$,
as well as the mass $M$ are shown as a function of
$\alpha^2=n_d/(8\kappa_1$) for regular solutions with $\tau_1=3\tau_2=1$
and two different values of $\tau$.
The mass of the flat space solution is $M=1.00236$ for $\tau=1$ and
$M=0.49257$ $(\tau=0.1)$}.
\\
\\
of the function $N(r)$
decrease, as indicated in Fig.~2.
These solutions exist up to a maximal
value $\alpha_{max}$ of the parameter $\alpha$,
which is smaller than the corresponding value in
the pure EYM theory~\cite{bcht}, and depends on the
value of
the coupling parameter $\tau$.
For example, we find numerically
$\alpha_{max}^2 \approx 0.2573$ fror $\tau=0.1$ while the
corresponding value for $\tau=1$ is $\alpha_{max}^2 \approx 0.06855$.
(Without a Grassmanian field, this branch extends up to
 $\alpha_{max}^2 \approx 0.5648$.)

Similar to the EYM case \cite{bcht}, we found another branch of solutions
on the interval $\alpha^2 \in [\alpha_{cr(1)}^2 , \alpha_{max}^2]$
with $\alpha_{cr(1)}^2 $ depending again on the value of $\tau$ 
(e.g. $\alpha_{cr(1)}^2  \approx 0.06302$ for $\tau=1$).
On this second branch of solutions, both $\sigma(0)$ and $N_m$ continue
to decrease but stay finite.
 However, a third branch of solutions exists
for $\alpha^2 \in [\alpha_{cr(1)}^2 , \alpha_{cr(2)}^2]$ ,
on which the two quantities decrease
further.
A fourth branch of solutions has also been
found, with a corresponding $\alpha_{cr(3)}^2$ close to
$\alpha_{cr(2)}^2$. Further branches of solutions, exhibiting more
oscillations very likely
exist but their study is a difficult numerical problem.
Progressing on this succession of branches, the main observation
is that the value $\sigma(0)$ decreases much faster than that of $N_m$
as illustrated in Fig.~2.  The pattern strongly suggests that after a
finite (or more likely infinite) number of oscillaions of $\sigma(0)$,
the solution terminates into a singular solution with $\sigma(0)=0$
and a finite value of $N(0)$.
As seen in Figure 2, the mass parameters do not increase
significantly along these secondary branches.

This is the behaviour observed in \cite{bcht} for the pure EYM theory.
The inclusion of a Grassmanian extra field does not seem to qualitatively
change the properties of the system.

In Fig.~3, we present the profiles of
the metric functions $N$ and $\sigma$  for the same value of
$\alpha^2$ on the first and second branch.
\newpage
\setlength{\unitlength}{1cm}
\begin{picture}(18,8)
\centering
\put(1,0){\epsfig{file=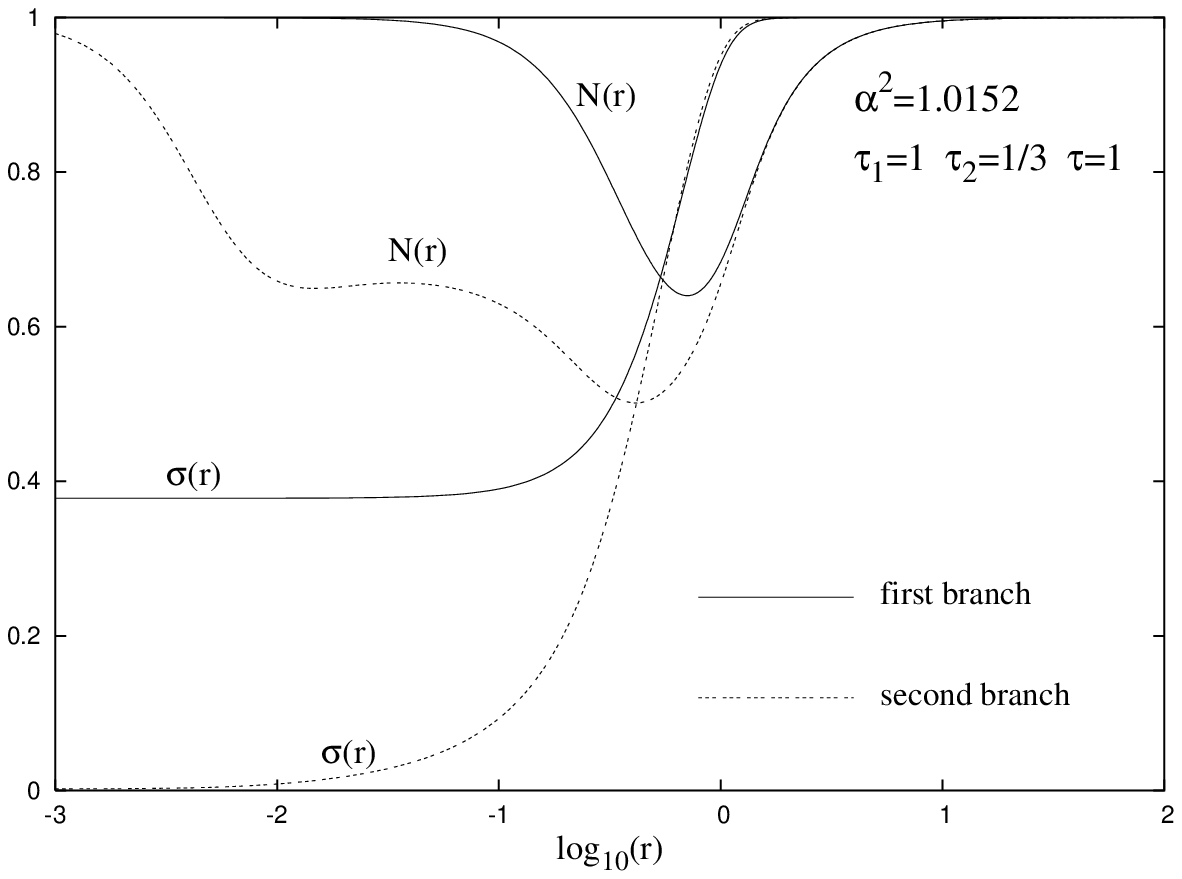,width=14cm}}
\end{picture}
\\
\\
{\small {\bf Figure 3.}
The metric functions $N(r),\sigma(r)$ are plotted as functions of radius
for typical regular gravitating solutions.}
\\
\\
The discussion in this section is restricted to the Grassmanian field
presenting no nodes. We do not expect the consideration of solutions with
nodes to change this picture.
\subsection{Black hole solutions}
\label{Blackhole}
According to the standard arguments, one can expect
black hole generalisations of the regular configurations to exist at
least for small values of the horizon radius $r_h$.
This is confirmed by the numerical analysis for solutions
with no nodes in $f(r)$ as well as solutions
with a frozen Grassmanian field $f(r)=f_{\infty}=0$.

Again, the properties of the solutions we find
are rather similar to the five dimensional
black hole solutions without a Grassmanian field
discussed in \cite{bcht}.
Firstly, black hole solutions seem to exist for all values of $\alpha$
for which regular solutions were constructed. Also, for a given set of
couplings ($\tau_1,\tau_2,\tau$), the solutions exist only
for a limited region of the $(r_h,\alpha)$ space.

The typical behaviour of solutions as function of $r_h$ is presented in
Figure 4, for a small value of $\alpha$ as compared to
to maximal value $\alpha_{max}$ of the regular solutions.
Starting from a regular solution and increasing the event horizon radius,
we find a first branch of solutions which extends to a maximal value
$r_{h(max)}$. The variation of mass and $\sigma(r_h)$ is relatively small
on this branch. Extending backwards in $r_h$, we find a second branch of
solutions for $r_h < r_{h(max)}$.
This second branch stops at some critical value $r_{h(cr)}$,
where the numerical iteration fails to converge.
The value of $\sigma(0)$ on this branch
decreases drastically, as shown in Fig.~4.
Also, the surface gravity $\kappa$ of the solutions,
given by
\begin{eqnarray}
\nonumber
\kappa^2=- \frac{1}{4}g^{tt}g^{rr}(\partial_r g_{tt})^2,
\end{eqnarray}
\newpage
\setlength{\unitlength}{1cm}
\begin{picture}(18,8)
\centering
\put(1,0){\epsfig{file=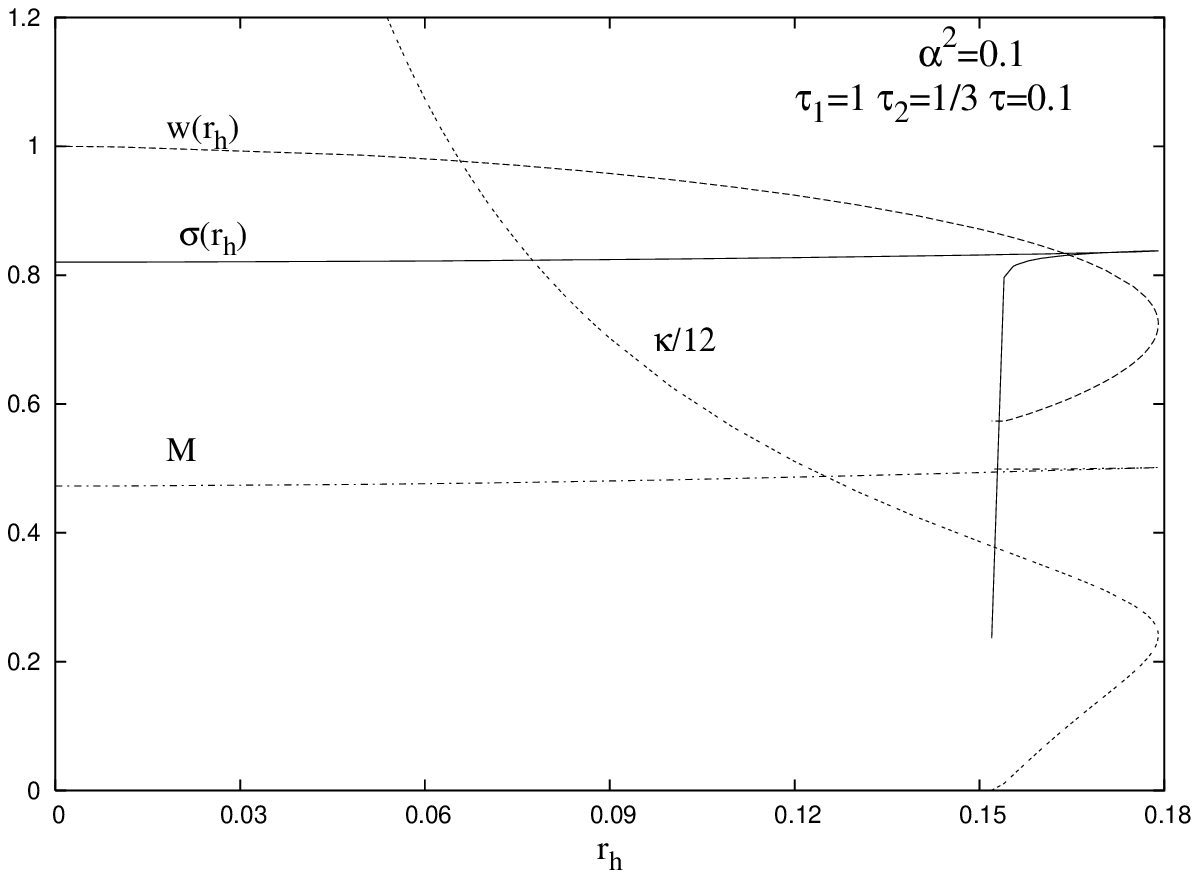,width=14cm}}
\end{picture}
\\
\\
{\small {\bf Figure 4.}
The value  $w_h$ of the gauge function at the horizon, the
value $\sigma_h$ of the metric function
$\sigma$ at the horizon, the mass $M$, as well
as the surface gravity $\kappa$ divided by 12 are
shown as functions of the event horizon radius $r_h$
for black hole solutions with $\alpha^2=0.1$,
$\tau_1=3\tau_2=1,\tau=0.1$.
These results are obtained for a "frozen" Grassmanian field $f=0$}.
\\
\\
strongly decreases on this
branch, approaching a very  small value.
However, the increase of the total mass is still very small.
Similar to the EYM case~\cite{bcht}, higher branches of solutions
on which the value $\sigma(0)$ continues to decrease further to zero
are likely to exist. However,
the extension of these branches in $r_h$ will be very small,
which makes their study difficult.

Although the results in Figure 4 correspond to a frozen Grassmanian field,
we do not expect a different result for solutions with a nontrivial
$f(r)$. Similar to the regular case, we find that the
node number of the
Grassmanian field does not significantly affect the propertiess of the
gravitating solutions.

However, the global picture  we find (and the corresponding EYM results)
may change by considering  values of $\alpha$ near $\alpha_{max}$.
We hope to come back on this point in a future work.

\section{Summary and discussion}
\label{conclusions}
The aim of this work is to find out whether a particular singular
behaviour of solutions of a gravitating YM model\footnote{The definition
of the model in question is $d$ (spacetime) dependent. While in all $d$
these are formally the same, nevertheless the gauge groups are
different, depending on $d$~\cite{bct2,bcht}.} in $d=5$ spacetime is a
persistent feature of the dimensionality of the spacetime? The reason for
asking this question is that the EYM model in question here
does not feature this critical behaviour in spacetimes $d=6,7,8$.

More simply stated, the model(s) in $d=6,7,8$ exhibit only a maximum
value of the gravitational coupling $\al^2=\al_{max}^2$, while that in
$d=5$ has, in addition to an $\al^2=\al_{max}^2$, also an
$\al_{cr}^2>0$. This is seen from the oscillatory behaviour of 
$\al^2$ converging to $\al_{cr}^2$ in Figure 2. Because $d=5$ spacetime
is particularly relevant in the context of the AdS/CFT correspondence,
and because the solutions in $d=5$ spacetime differ markedly in this
respect from those in $d=6,7,8$ spacetimes, it is important to study
this peculiar feature in $d=5$ further.

The original model(s) introduced in \cite{bct2} involve a dimensionful
constant in addition to the gravitational constant, analogously with the
$d=4$ gravitating YMH model~\cite{bfm,w}. Like the latter, regular
solutions to the higher dimensional EYM models exist only for values of
the gravitational coupling $\al^2$ up to a maximum $\al_{max}^2$. But this
analogy with the EYMH case goes further only in the $d=5$ EYM case, where
in addition to $\al_{max}^2$, there occurs also a critical $\al_{cr}^2>0$.
In the $d=6,7,8$ EYM cases it appears that $\al_{cr}^2=0$. This is
surprising since the $d=6,7,8$ EYM models support regular solutions in
the flat space limit just like the $d=4$ EYMH model, while the $d=5$ EYM
model does not support a flat space solution.

It is to throw some light on this question that we modified the $d=5$ EYM
model by introducing a (Grassmannian) scalar matter field, which results
in the new model supporting a flat space solution. The result is that
the qualitative features of the solutions of the original $d=5$ EYM
model are preserved.

In passing, we studied the (static) solitons of the $d=5$ flat space
$SU(2)$ gauged Grassmannian model, and found that these form an infinite
sequence of solutions exhibiting multi-nodes in the profile of the
Grassmannian profile function.

\medskip
\medskip

\noindent
{\bf\large Acknowledgements} This work is carried out
in the framework of Enterprise--Ireland Basic Science Research Project
SC/2003/390 of Enterprise-Ireland. One of us (YB) is grateful to the
Belgian FNRS for financial support.

\newpage

\begin{small}

\end{small}

\end{document}